\documentclass[reprint,
superscriptaddress,
 amsmath,amssymb,
 aps,
 prl,
]{revtex4-1}

\usepackage{graphicx}
\usepackage{dcolumn}
\usepackage{bm}
\usepackage{hyperref}
\usepackage{todonotes}
\usepackage{csquotes}
\begin{document}

\preprint{APS/123-QED}

\title{Atomtronic Matter-Wave Optics}

\author{Saurabh Pandey}
\affiliation{
 Institute of Electronic Structure and Laser, Foundation for Research\\
 and Technology-Hellas, Heraklion 70013, Greece
}
\affiliation{Department of Materials, Science and Technology, University of Crete,\\ Heraklion 70013, Greece}
\thanks{Los Alamos National Laboratory, Los Alamos, NM 87545, USA}

\author{Hector Mas}
\affiliation{
 Institute of Electronic Structure and Laser, Foundation for Research\\
 and Technology-Hellas, Heraklion 70013, Greece
}
\affiliation{Department of Physics, University of Crete, Heraklion 70013, Greece}
\thanks{Jet Propulsion Laboratory, California Institute of Technology, Pasadena, California 91109, USA}

\author{Georgios Vasilakis}
\affiliation{
 Institute of Electronic Structure and Laser, Foundation for Research\\
 and Technology-Hellas, Heraklion 70013, Greece
}

\author{Wolf von Klitzing}
\email{wvk@iesl.forth.gr}
\affiliation{
 Institute of Electronic Structure and Laser, Foundation for Research\\
 and Technology-Hellas, Heraklion 70013, Greece
}
 
\date{\today}% It is always \today, today,

\begin{abstract}
%Background
Matterwaves made up of ultra-cold 
quantum-degenerate atoms have enabled the creation of tools having unprecedented sensitivity and precision in measuring gravity, rotation or magnetic fields.
Applications range from gravitational wave detection and tests of Einstein's equivalence principle to inertial sensing for navigation and gravitational gradient sensing for oil and mineral exploration. 
In this letter, we introduce atom-optics as a novel tool of manipulating matterwaves   in ring-shaped coherent waveguides.
We collimate  and focus matterwaves   derived from Bose-Einstein Condensates (BECs) and ultra-cold thermal atoms in ring-shaped time-averaged adiabatic potentials.
We demonstrate `delta-kick cooling' of BECs, reducing their expansion energies by a factor of 34.
The atomtronic waveguide ring has a radius of only  $485\,\mu m$, compared to other state-of-the-art  experiments requiring zero gravity or chambers of ten meter. 
This level of control with extremely reduced spatial requirements is an important step towards atomtronic quantum sensors.

\end{abstract}

\flushbottom
\maketitle

Recent years have witnessed the rise of quantum technologies from pure \textit{Gedanken} experiments to real applications. 
One of the most striking examples is atom interferometry, which employs the wave-like nature of atoms to perform, amongst others, extremely sensitive measurements of inertial forces such as rotation \cite{Gustavson1997PRL} and acceleration \cite{Peters2001Metrologia}, making possible novel experiments in fundamental physics, e.g.~test of Einstein's equivalence principle \cite{ALTSCHUL2015ASR} and gravitational waves detection \cite{Dimopoulos2008PRD}.
On the applied side it enables inertial navigation, e.g.~in GPS denied environments, and the search for minerals and oil using gravitational mapping \cite{Angelis2008MSAT}.

Today state-of-the-art atom interferometers use freely falling atoms, where the interaction time between the atoms is limited by the height of the experimental apparatus, leading to impractical heights extending to 10\,m or even 100\,m \cite{Kasevich1991PRL, Giltner1995PRL, Zoest2010S,Muntinga2013PRL}.
Atomtronics  aims at miniaturising these atom-optical experiments and turn them into applied quantum technologies by manipulating  atoms in tiny, multiply-connected circuits---much like electrons in electronics or photons in waveguide circuits---and thus achieve much increased interrogation times and consquently much enhanced sensitivities in a much smaller space \cite{Wang2005PRL,McDonald2013PRA, Wu2007PRL, Gupta2005PRL}.
One of the key requirements of atomtronic interferometry is that the matterwaves have to be able to propagate along the waveguides without being disturbed.
There have been numerous attempts at producing such waveguides using optical and magnetic potentials, e.g.~in the study of the quantization of the superfluid flux \cite{Ryu2007PRL,Eckel2014Nat} in  ring-shaped magnetic \cite{Pandey2019N,Gupta2005PRL,Arnold2006PRA,Sherlock2011PRA} and optical \cite{Turpin2015OE,Bell2016NJP} potentials.
 The main challenge of decoherence-free propagation of matterwaves  in waveguides has been resolved recently using  time-averaged adiabatic potentials (TAAPs) as perfectly smooth waveguides, propagating condensates over distances of tens of centimeters at hypersonic speeds \cite{Lesanovsky2007PRL,Pandey2019N}.

\begin{figure*}
\centering
\includegraphics[width=0.75 \textwidth]{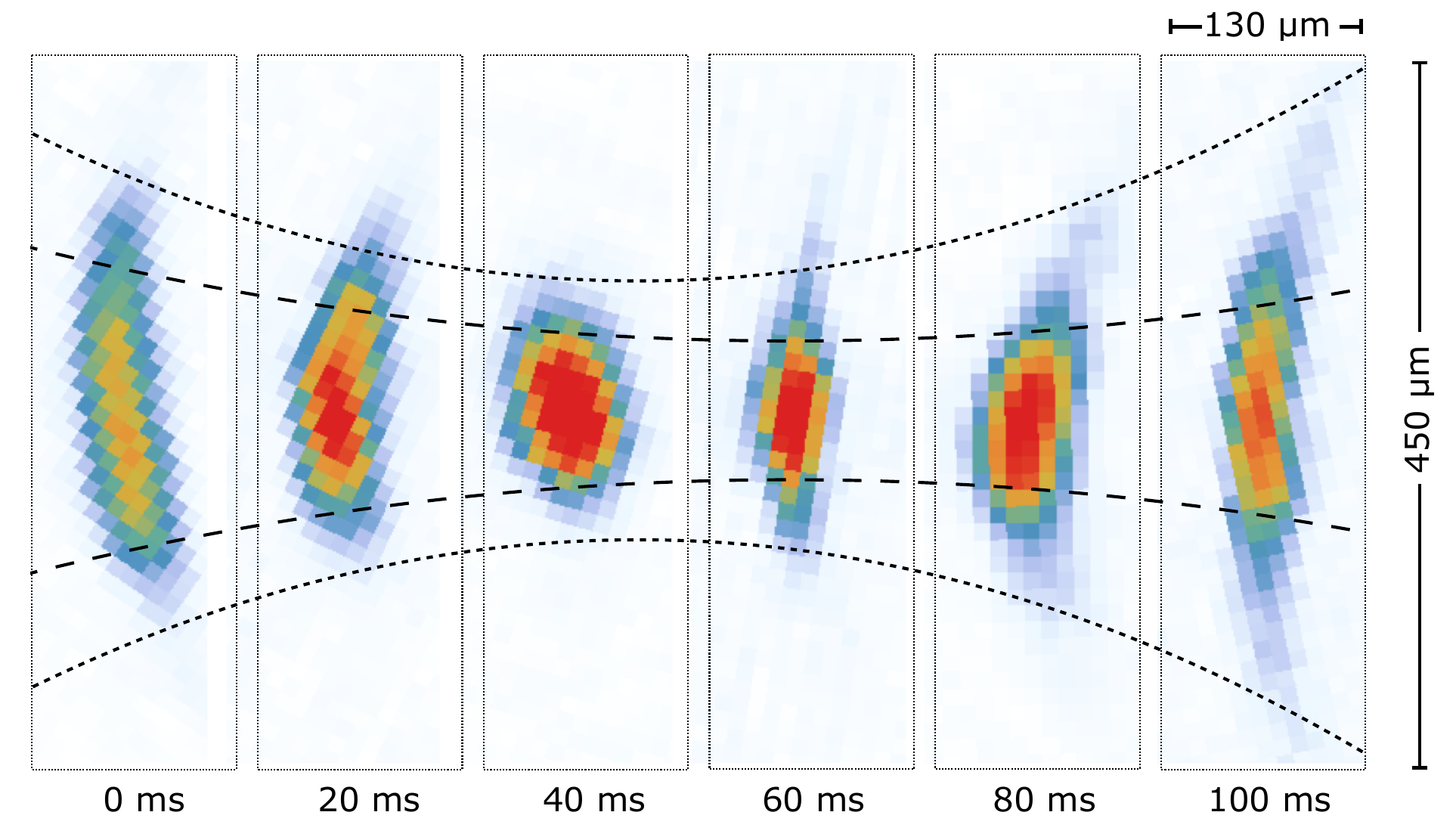}
\caption{Focusing of Bose Einstein Condensates and ultra cold thermal clouds in a ring-shaped matterwave guide. 
The absorption images were taken at different propagation times after the application of the matterwave  lens.
The long-dashed line is a guide to the eye for the BECs and the dotted line for the thermal clouds.
The images are taken after a time-of-flight of 5.3\,ms for a ring of radius $R=485\,\mu m$.
There were about $2\times 10^4$ atoms in the BEC and $6\times 10^4$ thermal atoms at a temperature of $166 \pm 33$\,nK. 
The vertical (horizontal) direction of the images is  tangential (radial) to the ring at the center of the condensate.
}
\label{fig:focusOD}
\end{figure*}

In this letter we address the remaining  key challenge in atomtronics: the management of the dispersion of the atomic wave packets during their propagation in the waveguide.
Just like in optical fibers, matterwave pulses travelling in the waveguides rapidly spread with propagation distance. 
For Bose-Einstein condensates this is caused by the self-interaction energy of the atoms and for thermal clouds it is due to the velocity spread of the atoms. 
If left unchecked, this makes high-sensitivity interferometry difficult to implement in an atomtronic device.

We introduce atomtronic matterwave  optics as a means to manipulate the density and momentum spread of Bose-Einstein Condensates and ultra-cold thermal atomic clouds in ultra-smooth TAAP waveguides.
We use a series of time-dependent gravito-magnetic matterwave lenses, which allow us to (de)focus or collimate matterwaves  in a waveguide (see Fig.\,\ref{fig:focusOD}).
We implement optimal control theory and delta-kick cooling/collimation as a means to manipulate the azimuthal momentum spread of BECs freely propagating in the ring-shaped waveguide.
Using an optimized matterwave lensing sequence we reduce the spread of BECs due to their atomic self-interaction by a factor of thirty four and the thermal spread of a thermal cloud by a factor of nine.

\begin{figure}[b]
\centering
\includegraphics[width=\columnwidth]{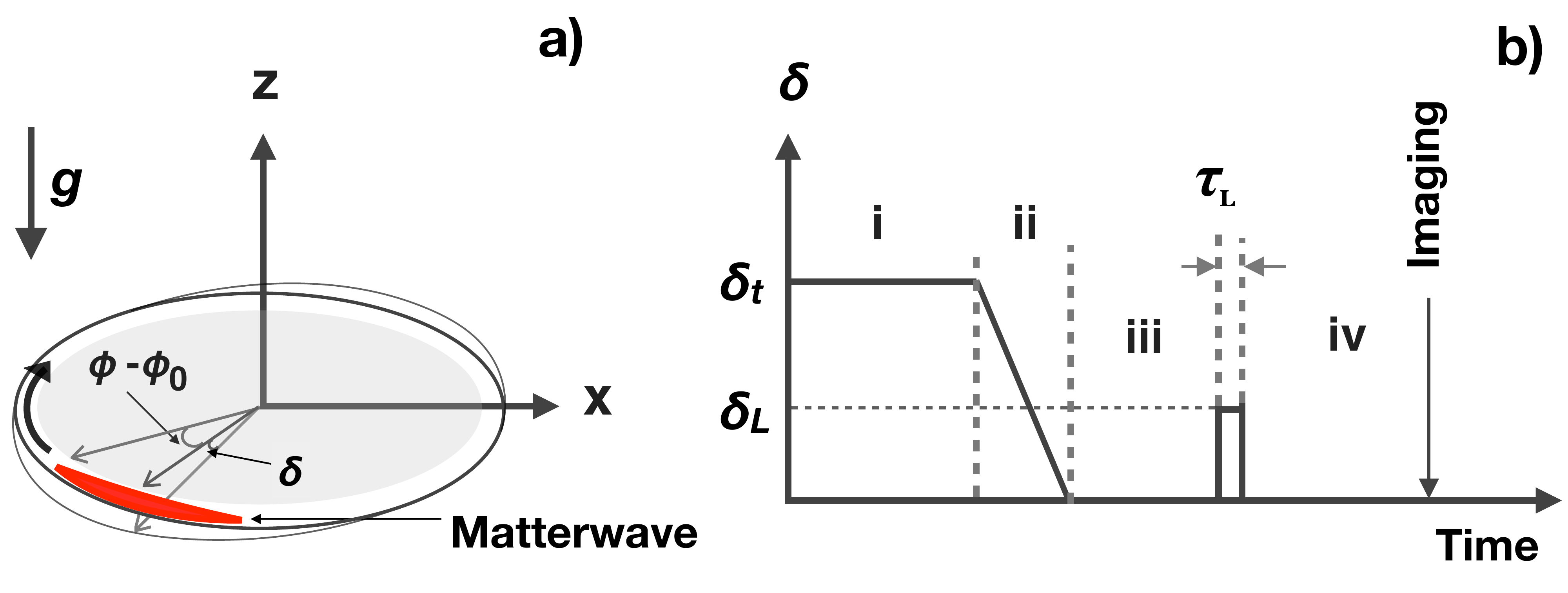}
\caption{a) Schematic of the matterwave guiding in the TAAP ring waveguide together with the azimuthal lens potential. 
b) Azimuthal tilt amplitude ($\delta$) for the atom optics sequence. (i) BEC acceleration, (ii)  adiabatic launch into the waveguide, (iii) a first free expansion of $\tau_\text{0}=66$\,ms in the waveguide, followed by a delta-kick pulse of duration $\tau_\text{L}=17$\,ms, and (iv) final variable expansion in the waveguide for a duration of $\tau_\text{F}$. 
This is followed by switching off all potentials and time of flight imaging.
The duration and tilt amplitudes at different stages are mentioned in the text.
}
\label{fig:schemeDKC}
\end{figure}

 We use ring-shaped waveguides based on Time-Averaged Adiabatic Potentials (TAAPs), which are formed using a combination of DC, audio- and radio-frequency (RF) magnetic fields \cite{Lesanovsky2007PRL}: 
The homogeneous RF field dresses the states of atoms in a DC magnetic quadrupole field. 
The ring-shaped potential results then from the time-averaging induced by an audio-frequency homogeneous magnetic field oscillating in the vertical direction.
TAAPs are extremely smooth with a residual roughness of well below 200\,pK \cite{Pandey2019N,Navez2016NJP}.
The gravito-magnetic matterwave lens results from tilting the waveguide against gravity.
The principle of atomtronic matterwave  optics using gravito-magnetic matterwave lenses
is outlined in Fig.\,\ref{fig:schemeDKC}: The atom cloud is loaded into a TAAP trap, accelerated, and released into the  ring-shaped waveguide, where it expands according to its temperature or interaction energy.
After some time a gravito-magnetic matterwave lens is applied for a duration $\tau_\text{L}$, which can either (de)focus or collimate the atomic cloud.
In atomtronic matterwave optics, just like in photon optics, one can collimate a point source by placing it at the focus of the lens either by varying the power of the lens or the distance at which it is placed. 
The optical equivalent of the  expansion time $(\tau_\text{0})$ is the object-lens distance. The strength of the parabolic lens potential corresponds to the curvature of the refractive index of a grin lens, and finally the time  $\tau_\text{L}$ corresponds to the thickness of the grin lens.  
The focal distance is then the second expansion time $\tau_\text{F}$.

%%%%%%%%%%%%%%%%%%%%%%%%%%%%%%%%%%%%%%%%%%%
\begin{figure*}[!ht]
\centering
\includegraphics[width=0.65 \textwidth]{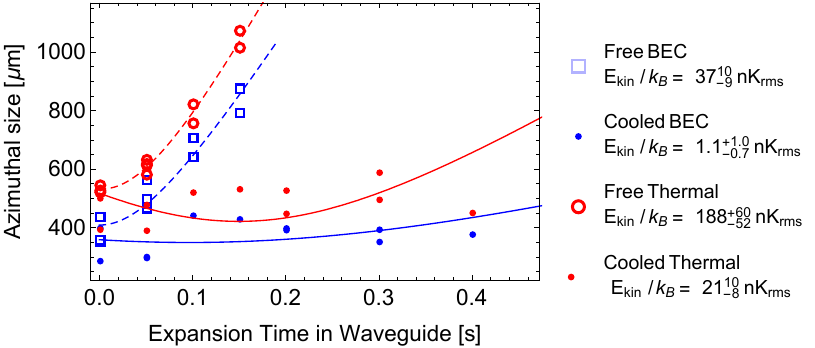}
\caption{ 
Collimation of guided BECs and thermal clouds in the ring waveguide.
The figure shows the azimuthal size of two-component atom clouds as they propagate freely in the ring-shaped matterwave  guide. 
The thermal size is the azimuthal $1/e$ 
radius $(R_{1/e})$  and for BECs the Thomas Fermi radius $(R_\text{TF})$. 
The dashed lines (and empty symbols) show the free expansion of a two component cloud in the ring-shaped waveguide in the absence of lensing.
The solid lines (and full symbols) show the free expansion of a two component cloud in the ring-shaped waveguide after near-optimal lensing.
The lines are a fit to  $\Delta l = R (\Delta \phi_0^2+\Delta\dot\phi^2\, (t-t_0)^2)^{1/2}$, with the average kinetic energy being $E_\text{kin,Therm}=1/2\,m(R_{1/e}\, \Delta\dot\phi )^2 $ and  $E_\text{kin,BEC}=1/7\,m(R_\text{TF}\, \Delta\dot\phi )^2 $.
The BECs contained about $1\times 10^{4}$ atoms and the thermal clouds $3\times 10^{4}$ atoms. 
The mean temperature measured by radial time-of-flight imaging is  225(39)\,nK for the cooled thermal cloud, 
and 181(8)\,nK for the freely expanded thermal cloud.
The cloud is travelling at an angular speed of 31\,mm/s in a circular waveguide of length 3.1\,mm, i.e. the figure spans more than four round trips in the waveguide. 
}
\label{fig:azimuthalSizeThermalAndBEC}
\end{figure*}
%%%%%%%%%%%%%%%%%%%%%%%%%%%%%%%%%%%%%%%%%%%

We begin by studying the free expansion of thermal clouds and BECs in the ring-shaped TAAP waveguide.
We  load  atoms from a dipole trap into the ring-shaped waveguide with the superimposed azimuthal trapping potential similar to the one that later forms the lens.
We do so without inducing any measurable excitation of a center-of-mass or shape oscillation. 
We accelerate ultra-cold thermal clouds and  BECs to 31\,mm/s using the bang-bang control sequence of optimal control theory \cite{Chen2011PRA, Pandey2019N} and launch them into the TAAP waveguide simply by lowering the azimuthal trapping potential.
The azimuthal size of the cloud then increases at a rate determined by its temperature (or by the chemical potential for BECs) until the atoms fill the whole ring. 
After the atom-optical manipulation and some propagation in the waveguide we switch off the guiding potential, let the atom clouds expand freely and image them (see Fig.\,\ref{fig:focusOD}). 
The size of atomic clouds then is a direct measure of the expansion energy due to  temperature or chemical potential, which can be expressed as an effective temperature  $T_\text{rms}=m v_\text{rms}^2/k_\text{B}$, where $m$ is the mass of one atom, $v_\text{rms}$ the rms velocity of the atoms, and $k_\text{B}$ Boltzmann's constant \cite{Kovachy2015PRL}.
The empty symbols in Fig.\,\ref{fig:azimuthalSizeThermalAndBEC}  show the free expansion of BECs (squares) and ultra-cold thermal clouds  (circles) in the waveguide.
From the fit of the expansion we deduce an azimuthal
kinetic energy of $T_\text{rms}=188^{+60}_{-52}$\,nK for the thermal cloud and $37$\,nK for the BEC.
This agrees nicely with the mean temperature measured from time-of-flight expansion in the radial direction ($T_\text{rad}=181\pm8$\,nK).

%%%%%%%%%%%%%%%%%%%%%%%%%%%%%%%%%%%%%%%%%%%
\begin{figure}[b]
\centering
\includegraphics[width=0.95\columnwidth] {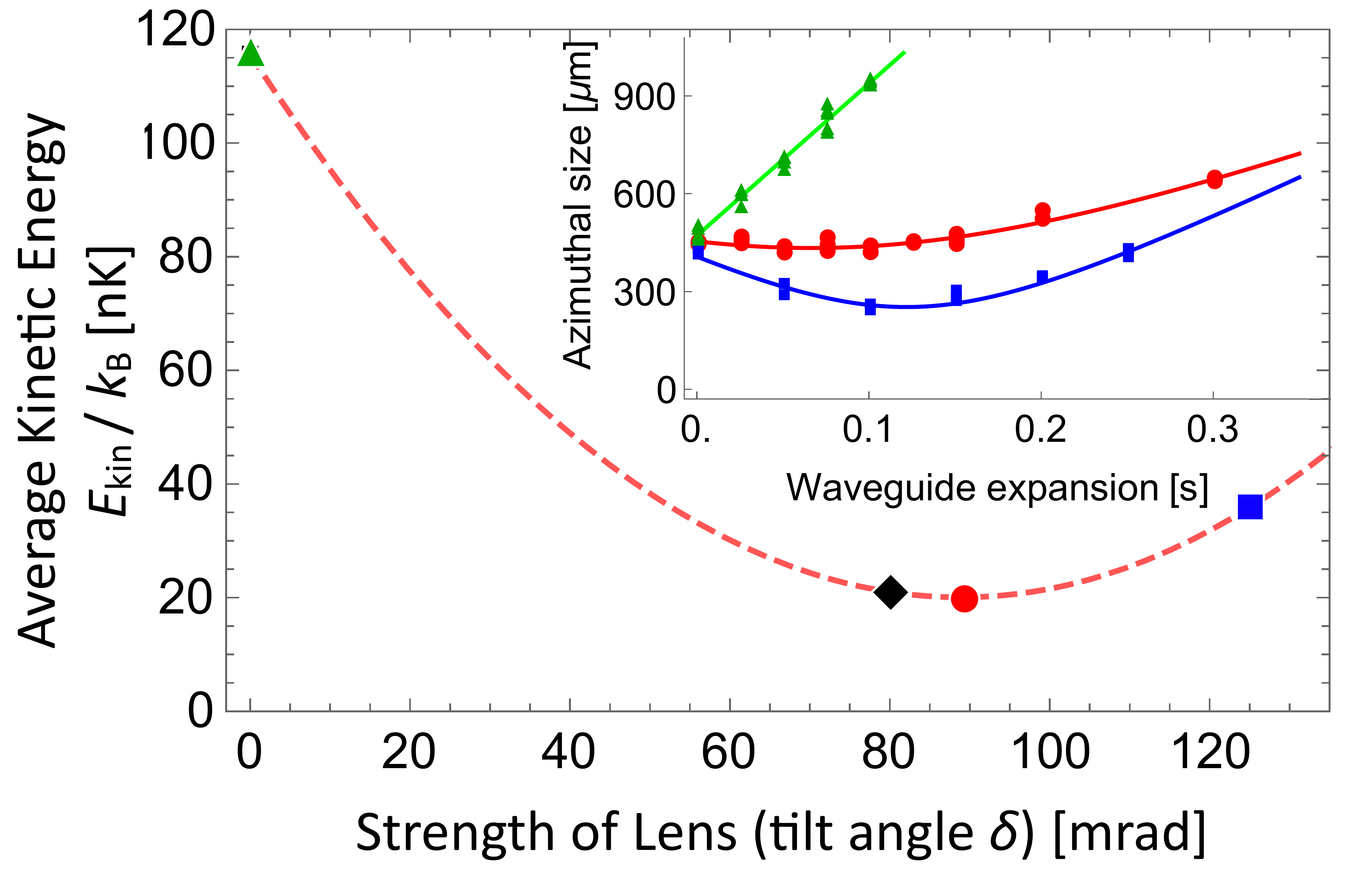}
\caption{
The main panel shows the kinetic energy of a thermal cloud as a function of the strength of the focusing lens (the tilt angle $\delta$ of the ring-shaped waveguide). 
The insert shows the expansion along the waveguide for the different strengths of the lenses: no lensing ($\delta=0$, green triangles), optimal collimation ($\delta=89$\,mrad, red dots), and focusing  ($\delta=125$\,mrad, blue squares). 
The thermal clouds originally contain $6\times10^{4}$ atoms at $116^{+5}_{-5}$\,nK.
The lowest kinetic energy after atomtronic lensing is $21^{+2}_{-1}$\,nK, corresponding to a `delta-kick cooling' by a factor of six. 
  }
\label{fig:expansionSlopes}
\end{figure}
%%%%%%%%%%%%%%%%%%%%%%%%%%%%%%%%%%%%%%%%%%%

Having studied free expansion of thermal and quantum degenerate clouds, we now turn to atomtronic matterwave optics.  
The gravito-magnetic matterwave lenses, as laid out in Fig.\,\ref{fig:schemeDKC} and described above, consist of an azimuthal harmonic potential co-moving with the atomic cloud.
We adjust this matterwave optical lens system
by varying the tilt $\delta$ of the ring and thus the strength of the co-moving potential and  its focal length.
We then determine the expansion velocities by observing the size of the a thermal cloud after some time of free propagation.
Just like in an optical system, a focus is achieved for the smallest (cloud) size and collimation for the smallest beam divergence or expansion energies.
The insert in Fig.\,\ref{fig:expansionSlopes} shows the  expansion  of the atomic clouds for different strengths of the gravito-magnetic lens.
The main  panel shows the expansion energies as a function of the strength of the lens. 
Starting from a free expansion energy of $116(5)\,\text{nK}$, the lowest kinetic energy ($21^{+2}_{-1}$\,nK) is achieved for a tilt of  $\delta=89\, \text{mrad}$. 
The lensing has therefore reduced the initial temperature of the cloud in the azimuthal direction by a factor of six in a waveguide of only $485\,\mu m$ radius.

In close analogy to photon waves, the focusing of matterwaves  is limited by spatial coherence with the thermal cloud closely resembling the radiation from an incandescent light bulb and the Bose-Einstein Condensates being the equivalent of lasers.
Fig.\,\ref{fig:azimuthalSizeThermalAndBEC} shows the azimuthal evolution of  thermal and Bose condensed atom  clouds traveling together around the ring waveguide, either with (solid symbols) or without (open symbols) the matterwave lens.
A fit of the azimuthal expansion velocities of the freely expanding clouds reveals average kinetic energies of $188^{+60}_{-52}$\,nK for the thermal cloud and $37^{+10}_{-9}$\,nK for the BEC. 
We focus both using a matterwave lens of a duration of $\tau_{\text{L}}=17$\,ms and tilt $\delta$ = 70\,mrad.
The solid symbols in Fig.\,\ref{fig:azimuthalSizeThermalAndBEC} show the expansion in the waveguide of BEC and thermal cloud after the pulse is turned off. 
A fit reveals  kinetic temperature energies of $1.1^{+1.0}_{-0.7}$\,nK  for the BEC and $21^{+10}_{-8}$\,nK for the thermal cloud.
Comparing with the no-kick case, there is a reduction in the kinetic energy (`delta-kick cooling')  by a factor of 34 for the BECs and a factor of 9 for the thermal clouds.
This has to be compared to current state of the art, where Kovachy et al.~\cite{Kovachy2015PRL} demonstrated a reduction in energy by a factor of 32 using a 10\,m vacuum chamber and to M\"untinga et al.~\cite{Muntinga2013PRL}, who also achieved a reduction of the condensate's kinetic energy to 1\,nk, albeit in a 100\,m drop tower.

In conclusion, we have demonstrated for the first time an atomtronic device capable of atom optical manipulation of BECs and thermal clouds in a ring-shaped waveguide.  
The achieved reduction in temperature and its final value rival the best realized so far.
The reduction of the spatial requirements by more than four orders of magnitude compared to state-of-the-art atom-optical experiments together with the extreme precision of the atom-optical manipulations open the field of atomtronic  devices to coherent precision manipulation of matterwaves.
Future applications will include atom-interferometry, ultra-low collisions and precision 1D/3D BEC physics.

%%%%%%%%%%%%%%%%%%%%%%%%%%%%%%%
\subsection{Acknowledgements}
%%%%%%%%%%%%%%%%%%%%%%%%%%%%%%%
This work has received funding from the European Union's Horizon 2020 research and innovation programme H2020-FETOPEN-2018-2019-2020-01 under grant agreement No 863127  \enquote{nanoLace}.
It was also supported by the project \enquote{HELLAS-CH} (MIS 5002735) which is implemented under the \enquote{Action for Strengthening Research and Innovation Infrastructures},
funded by the Operational Programme \enquote{Competitiveness, Entrepreneurship and Innovation} (NSRF 2014-2020) and co-financed by Greece and the European Union (European Regional Development Fund). 
GV received funding from the European Union’s Horizon 2020 research and innovation programme under the Marie Skłodowska-Curie Grant Agreement No  750017.
SP acknowledge financial support from the
Hellenic Foundation for Research and Innovation (HFRI) and the General Secretariat and Technology (GSRT), under the HFRI PhD Fellowship grant (4823).

\subsection{Author contributions}
WK conceived the experiments. 
SP, WK and HM designed and built the experiment, and
SP carried them out. 
WK and SP analysed the results. 
SP and WK wrote the manuscript. 
All authors edited and reviewed the manuscript.

%%%%%%%%%%%%%%%%%%%%%%%%%%%%%%%
\subsection{Conflict of Interest}
%%%%%%%%%%%%%%%%%%%%%%%%%%%%%%%
The authors declare that they have no conflict of interest.

%%%%%%%%%%%%%%%%%%%%%%%%%%%%%%%
%\bibliography{atomtronics.bib}
%%%%%%%%%%%%%%%%%%%%%%%%%%%%%%%

%

\end{document}